\documentclass[12pt]{article}
\usepackage{epsfig}
\usepackage{psfrag}
\usepackage{amsmath}
\usepackage{lscape}
\usepackage{multirow}
\usepackage[super, comma, sort&compress]{natbib}
\newcommand{\bfig}{\begin{figure}[h] \begin{center}}
\newcommand{\efig}{\end{center} \end{figure}}

\def\longrightharpoonup{\relbar\joinrel\rightharpoonup}
\def\longleftharpoondown{\leftharpoondown\joinrel\relbar}

\def\longrightleftharpoons{
  \mathop{
    \vcenter{
      \hbox{
      \ooalign{
        \raise1pt\hbox{$\longrightharpoonup\joinrel$}\crcr
	  \lower1pt\hbox{$\longleftharpoondown\joinrel$}
	  }
      }
    }
  }
}

\newcommand{\rrxntwoone}[4]{\ensuremath{\displaystyle {\mathrm{#1}+\mathrm{#2}\longrightleftharpoons_{k^-_{#4}}^{k^+_{#4}}\mathrm{#3}}}}

\newcommand{\rrxntwotwo}[5]{\ensuremath{\displaystyle {\mathrm{#1}+\mathrm{#2}\longrightleftharpoons_{k^-_{#5}}^{k^+_{#5}}\mathrm{#3}+\mathrm{#4}}}}

\newcommand{\irxntwotwo}[5]{\ensuremath{\mathrm{#1}+\mathrm{#2}\stackrel{\kappa_{#5}}{\longrightarrow}}\mathrm{#3}+\mathrm{#4}}

\newcommand{\irxnonetwo}[4]{\ensuremath{\mathrm{#1}\stackrel{\kappa_{#4}}{\longrightarrow}}\mathrm{#2}+\mathrm{#3}}

\newcommand{\rrxnoneone}[3]{\ensuremath{\displaystyle {\mathrm{#1}\longrightleftharpoons_{k^-_{#3}}^{k^+_{#3}}\mathrm{#2}}}}

\newcommand{\mmrxn}[6]{\ensuremath{\mathrm{#1}+\mathrm{#2}\longrightleftharpoons_{k^-_{#6}}^{k^+_{#6}}\mathrm{#3}\stackrel{\kappa_{#6}}\longrightarrow\mathrm{#4}+\mathrm{#5}}}

\newcommand{\irxnoneone}[3]{\ensuremath{\displaystyle {\mathrm{#1}\stackrel{\kappa_{#3}}\longrightarrow\mathrm{#2}}}}

\begin{document}

\begin{center}
{\large Stochastic modeling of p53-regulated apoptosis upon radiation
damage}
\end{center}
\begin{center}
Divesh Bhatt, Zoltan Oltvai, and Ivet Bahar.
\end{center}

\section{Introduction}

 Studying cellular response to radiation damage is important from
the perspectives of both radiotherapy and the mitigation of
radiation damage. In the case of the former, the goal
is the induction of cell death, whereas in the latter case a
cellular response leading to organismal survival is required.

 The fate of the organism after radiation damage is linked
in a complex manner to cellular fate. In the short term,
an extensive cell death due to radiation damage will lead to
the death of the organism. In the long term (relevant only if the organism
survives the short term effects), it the survival of healthy and robust cells
that dictate the organism's survival (for example, absence of long term
cancerous cells). In this manuscript, we focus on the immediate
survival of the organism after radiation damage.

 The mechanism of cell death after radiation damage is purported to
be apoptosis via caspase activation that occurs several hours after
radiation injury. Thus, it is natural to investigate cellular
apoptotic machinery. Cellular response to radiation is
very complex and involves, possibly, proteins that respond to DNA damage
and the formation of free radicals that modify cellular biochemistry.
Although the exposure to radiation may be momentary, the effect of
the exposure on cellular biochemistry may be long lived. Further,
several proteins that are expressed transiently after radiation
damage, may trigger downstream response such as caspase activity
that occurs long after their expression.

 In view of the complex cellular response, protein--protein interactions
that lead to that response, and possible importance of temporal
evolution of the system, it is important to study systemically the 
time-dependent cellular response to stress.

 There have been various efforts to model cellular response to radiation,
as well as to model apoptosis. Efforts to model cellular response
to radiation were largely precipitated by the observation, at
the level of a single cell, of oscillatory p53 response to 
radiation damage.\cite{Or00}
In particular, the oscillatory response of p53 to
radiation damage has been modeled mostly using deterministic simulations 
with kinetic rate laws,\cite{Or00,Zatorsky06,Ma05,Loewer10}
and, to a significantly lesser extent, via stochastic
simulations.\cite{Proctor08}
Cellular apoptosis
has been modeled, independent of p53 radiation response, 
via deterministic simulations by several research groups,
\cite{Eissing04,zerrin1,Legewie06,Albeck08}
including efforts to include cell--cell variability in a
probabilistic manner.\cite{Spencer09,Skommer10}

Recently, there have been a few efforts to link the p53 response
to cell survival/death using deterministic 
simulations,\cite{Wee09,Pu10,Li11} or with inclusion of
limited stochasticity.\cite{Zhang09}
Due to increasingly larger pool of experimental observation on
the proteins involved, and their cellular compartmentalization, 
in the apoptotic network, modeling of transient system behavior
remains an active area of research and serves as a platform for
evaluating pharmacological strategies.

 Spatio-temporal modeling using stochastic simulation methods
allows for detailed examination of the kinetics of biochemical reaction
networks within a cell. From such simulations, one can address several 
different issues: spatial localization of reactions ({\it e.g.}, on the
mitochondrial membrane), as well as studying the time evolution of
molecular concentrations in response to perturbations to the system.
These perturbations can be radiation stress on the
cell, or the response to a given drug dose. Further, stochasticity
allows for the inclusion of cell variability in a natural way,
as well as for
studying conditions where the number of certain types of molecules
become very low and their presence/absence dictates system behavior.

In this study, we perform stochastic simulations to study the time
evolution of proteins, associated with apoptotic pathways,
in response to radiation damage. Further, we check the
efficacy of various treatment strategies, including
polypharmacological effects. In particular, we
use different inhibitors targeting different
proteins in the reaction network.
The manuscript is organized as follows. We first present the
biochemical reaction network, gathered from experimental
observations, associated with radiation induced apoptosis.
Then, we describe briefly the stochastic simulations that we perform
to study cell response to radiation damage, followed by results and
discussion.

\section{Biochemical reaction network}

 We present here the chain of biochemical events that we use to
model apoptotic mechanisms that are activated after radiation damage.
Instead of adopting a comprehensive model that includes all the 
cellular processes that are activated, or affected, by
radiation damage, we focus on a number of key proteins/genes
as a first approximation. An overview of the key interactions
is given in Figure~\ref{f0}. Table~\ref{t1}
gives details about the proteins involved in the system.
Several other proteins besides those in Figure~\ref{f0} are 
included in the model and described in Table~\ref{t1}, since
an interaction map, such as that in Figure~\ref{f0}, does not
contain details of the biochemical events that lead to the
individual interactions.

 Before we describe the radiation response of the network, we present
the generic underlying network.

\subsection{p53 module}

 p53 is an integral component of cell death due to genotoxic stress
(including radiation damage), and it activates apoptosis via both
transcription--dependent and 
transcription--independent pathways.\cite{Haupt95,Caelles94,Chipuk04}
The exact effect of p53 in its transcription-independent role in
apoptosis is still in-debate (see a recent review by Lindenboim {\it et
al.}\cite{Lindenboim11}), we attempt to incorporate some of the
key experimental observations.

The response of p53 to radiation exposure has been studied in great
detail at the individual cell level. The response of p53 to
radiation determined experimentally forms the starting point of
our modeling effort:
we first focus on the
p53 ``module'' (as highlighted in Figure~\ref{f0} with the dashed
rectangle) and calibrate our model to reproduce the known p53 response.

 p53 exists in several forms in a cell based on its post-translational 
modifications ({\it e.g.}, phosphorylation, ubiquitination at different sites).
In this manuscript, we utilize a model that
captures the known biological features: such as p53 oscillatory behavior
upon radiation damage,\cite{Zatorsky06,Or00,Lahav04}
and a stress--induced increase in nuclear p53\cite{Marchenko10} and its
transcriptional activity.\cite{Joers04,Lee10a}

p53 is translated in the cytoplasm (p53C) and this formation is
represented by
\begin{equation}
\irxnoneone{\Phi}{p53C}{1}
\label{e1}
\end{equation}
where $\Phi$ represents a null state, and the above reaction represents
the formation of p53C under normal operating conditions 
given by the rate $\kappa_1$.

 p53C can translocate to the nucleus, especially when a cell undergoes
stress.\cite{Marchenko10}
This translocation is modeled by
\begin{equation}
\irxnoneone{p53C}{p53n}{2}
\label{e2}
\end{equation}
where p53n is the nuclear p53. In the nucleus, p53n transcriptionally
activates Mdm2 mRNA and this transcription activation is modeled as
cooperative:\cite{Ma05} p53n tetramers on the DNA leads to this transcription
activation. The oligomerization of p53n is depicted by
\begin{equation}
\rrxnoneone{4p53n}{p53no}{3}
\label{e3}
\end{equation}
where the reaction order of the forward reaction is 4.
The oligomerized p53 leads to a direct transcription activation
of Mdm2:
\begin{equation}
\irxnonetwo{p53no}{p53no}{Mdm2m}{4}
\label{e4}
\end{equation}

 The Mdm2 messenger RNA translocates to the cytoplasm where it is
translated. The cytoplasmic Mdm2 (Mdm2C) can be imported into
the nucleus. This series of events is described by the following reactions:
\begin{equation}
\irxnoneone{Mdm2m}{Mdm2mC}{5}
\label{e5}
\end{equation}
\begin{equation}
\irxnoneone{Mdm2mC}{Mdm2C}{6}
\label{e6}
\end{equation}
\begin{equation}
\irxnoneone{Mdm2C}{Mdm2}{7}
\label{e7}
\end{equation}

 The monoubiquitinated, nuclear p53 (p53NU) is formed via the interaction of
p53n with Mdm2 (or similar ligases),\cite{Lee10} and we represent 
this by
\begin{equation}
\mmrxn{p53n}{Mmd2}{p53.Mdm2}{p53NU}{Mdm2}{8}
\label{e8}
\end{equation}
A similar reaction is modeled to occur for the cytoplasmic p53 and Mdm2:
\begin{equation}
\mmrxn{p53C}{Mmd2C}{p53C.Mdm2}{p53CU}{Mdm2C}{9}
\label{e9}
\end{equation}
Further, the ubiquitinated nuclear p53 can translocate to the 
cytoplasm:
\begin{equation}
\irxnoneone{p53NU}{p53CU}{10}
\label{e10}
\end{equation}
The ubiquitinated cytoplasmic p53 can translocate to the 
mitochondria\cite{Marchenko07}
as shown below:
\begin{equation}
\irxnoneone{p53CU}{p53m}{11}
\label{e11}
\end{equation}

 Upon severe genotoxic stress, p53n become transcriptionally active
for pro-apoptotic proteins\cite{Lee10a} such as PUMA and Bax, and 
such a transcriptionally active form is represented by p53T:
\begin{equation}
\irxnoneone{p53n}{p53T}{12}
\label{e12}
\end{equation}
Although the major role of p53T is transcriptional activation of
pro-apoptotic proteins, we let it retain the capacity for
oligomerizing and activating Mdm2 (this, though, is a minor effect),
\begin{equation}
\rrxnoneone{4p53T}{p53To}{13}
\label{e13}
\end{equation}
\begin{equation}
\irxnonetwo{p53To}{p53To}{Mdm2}{14}
\label{e14}
\end{equation}

The increase in
p53 upon radiation damage is not expected to be a step function. Similarly,
return of p53 to baseline levels after removal of radiation source is 
not expected to be instantaneous. Indeed, oscillatory levels of p53
persist for an extended period of time after radiation 
damage.\cite{Zatorsky06}
Further, p53 response to radiation is mediated by upstream
proteins such as ATM.
We do not model such upstream agents in specific detail, instead
we represent the upstream events by the following set of reactions.
\begin{equation}
\rrxnoneone{\Phi}{A}{15}
\label{e15}
\end{equation}
\begin{equation}
\irxnonetwo{A}{A}{B}{16}
\label{e16}
\end{equation}
\begin{equation}
\irxnoneone{B}{\Phi}{17}
\label{e17}
\end{equation}
\begin{equation}
\irxnonetwo{B}{B}{p53C}{18}
\label{e18}
\end{equation}
Thus, A activates B, which in turn activates p53C. The effect of these
upstream reactions is a smooth increase in p53C concentration
after radiation damage.

 In addition to the above equations, we model the independent degradation
of p53C, p53n, p53CU, p53NU, p53T, and p53m to highlight the fact that
external agents ({\it i.e.}, those not modeled explicitly) have a role 
in modulating protein concentrations. For example, several E3-ligases
(other than Mdm2) ubiquitinate and label p53 for proteasomal degradation
downstream. These reactions are given below:
\begin{equation}
\irxnoneone{p53C}{\Phi}{19}
\label{e19}
\end{equation}
\begin{equation}
\irxnoneone{p53n}{\Phi}{20}
\label{e20}
\end{equation}
\begin{equation}
\irxnoneone{p53CU}{\Phi}{21}
\label{e21}
\end{equation}
\begin{equation}
\irxnoneone{p53NU}{\Phi}{22}
\label{e22}
\end{equation}
\begin{equation}
\irxnoneone{p53T}{\Phi}{23}
\label{e23}
\end{equation}
\begin{equation}
\irxnoneone{p53m}{\Phi}{24}
\label{e24}
\end{equation}

\subsection{p53 modulated apoptotic pathways}

 Now, we discuss the mitochondrial apoptotic pathways downstream of
p53, and the role of p53 module to the activation of mitochondrial
apoptosis (see Figure~\ref{f0}).

p53T leads to transcription activation of proapoptotic PUMA and Bax, as
modeled by
\begin{equation}
\irxnonetwo{p53T}{p53T}{PUMA}{25}
\label{e25}
\end{equation}
and,
\begin{equation}
\irxnonetwo{p53T}{p53T}{Bax}{26}
\label{e26}
\end{equation}
The roles of PUMA and Bax are discussed below.

p53m binds to the anti-apoptotic protein Bcl-2 to 
inhibit the anti-apoptotic action of Bcl-2 on the mitochondria.\cite{Tomita06} 
The biochemical equation associated with this reaction is given by
\begin{equation}
\rrxntwoone{p53m}{\text{Bcl-2}}{p53m.\text{Bcl-2}}{27}
\label{e27}
\end{equation}
with binding affinity given by $k_{27}^+/k_{27}^-$.
PUMA also binds to the antiapoptotic Bcl-2 member on the mitochondrial
membrane.\cite{Chipuk05}
\begin{equation}
\rrxntwoone{PUMA}{\text{Bcl-2}}{PUMA.\text{Bcl-2}}{28}
\label{e28}
\end{equation}
Further, PUMA binds stronger to Bcl-2 than does p53, and can
displace p53 from its complex with Bcl-2\cite{Chipuk04} - as can
be seen by a combination of reactions~\ref{e27} and \ref{e28},
thus, freeing up p53 for
further proapoptotic activity on the mitochondria, such as
by activating Bax/Bak as shown below.

Pro-apoptotic Bax translocates back and forth from the cytoplasm to
the mitochondria\cite{Edlich11} to establish an equilibrium between
the cytosolic and mitochondrial Bax concentrations:
\begin{equation}
\rrxnoneone{Bax}{Bax_{mito}}{29}
\label{e29}
\end{equation}
At the mitochondria, Bax interacts with Bcl-2 to form a 
complex\cite{Edlich11} that prevents Bax oligomerization:
\begin{equation}
\rrxntwoone{Bax_{mito}}{\text{Bcl-2}}{Bax_{mito}.\text{Bcl-2}}{30}
\label{e30}
\end{equation}
Thus, the available Bcl-2 (that also forms complexes with p53 and PUMA 
as shown above) regulates the amount of Bax available for oligomerization.

Chipuk {\it et al.}\cite{Chipuk05} reported that
p53m binds stronger to Bcl-2 than does Bax, and can displace Bax from
its complex with Bcl-2. This reaction can be modeled via the following
reaction:
\begin{equation}
\rrxntwotwo{p53m}{Bax_{mito}.\text{Bcl-2}}{p53m.\text{Bcl-2}}{Bax_{mito}}{31}
\label{e31}
\end{equation}
However, eq~\ref{e31} is a straightforward linear combination of
eqs~\ref{e27} and \ref{e30}, and, thus, the obvious requirement of
consistent equilibrium constants
$(k_{31}^+/k_{31}^- = k_{27}^+k_{30}^-/k_{37}^-k_{30}^+)$ must be
observed (or, eq~\ref{e31} should not be considered explicitly).

 Active form of Bax on the mitochondrial membrane is needed for
oligomerization, a process that requires further interaction of
Bax with activators such as tBid. 
tBid localization to 
mitochondria\cite{Lutter01,Gonzalvez05} is an important event for Bax
activation and is modeled by the following reaction:
\begin{equation}
\rrxnoneone{tBid}{tBid_{mito}}{32}
\label{e32}
\end{equation}
Subsequently, tBid activates mitochondrial Bax into an active (for
oligomerization) form:
\begin{equation}
\irxntwotwo{tBid_{mito}}{Bax_{mito}}{tBid_{mito}}{Bax^*}{33}
\label{e33}
\end{equation}
It has also been suggested that tBid activates Bax in the cytosol
which then localizes to the mitochondria.\cite{Eskes00} However,
the order of consecutive reactions is not expected to have
a significant effect further downstream (such as caspase activation).
A similar Bax (or Bak) activating effect can be achieved by p53m:\cite{Chipuk04}
\begin{equation}
\irxntwotwo{p53m}{Bax_{mito}}{p53m}{Bax^*}{34}
\label{e34}
\end{equation}
and by PUMA,\cite{Ren10}
\begin{equation}
\irxntwotwo{PUMA}{Bax_{mito}}{PUMA}{Bax^*}{35}
\label{e35}
\end{equation}

The activated Bax then oligomerizes on the mitochondrial membrane:
\begin{equation}
\rrxntwoone{Bax^*}{Bax^*}{Bax_2}{36}
\label{e36}
\end{equation}
Further, higher order oligomerization is modeled following Albeck
{\it et al.}\cite{Albeck08}
\begin{equation}
\rrxntwoone{Bax_2}{Bax_2}{Bax_4}{37}
\label{e37}
\end{equation}
Mitochondrial outer membrane permeabilization
(MOMP) induced by Bax oligomerization results in the release of
cytochrome-{\it c} from the mitochondria into the cytoplasm.
We model the release of cytochrome-{\it c} via the following 
process:\cite{zerrin1}
\begin{equation}
\irxntwotwo{Bax_4}{cyt{\it c}_{mito}}{Bax_4}{cyt{\it c}}{38}
\label{e38}
\end{equation}
where the above reaction serves to model the fact that the presence
of a Bax pore is essential for release of cytochrome-{\it c} into
the cytoplasm.
Similarly, MOMP also releases Smac/Diablo into the cytoplasm: these proteins
bind to anti-apoptotic IAP's and, thus, facilitate apoptosis.\cite{Du00}
\begin{equation}
\irxntwotwo{Bax_4}{Smac_{mito}}{Bax_4}{Smac}{39}
\label{e39}
\end{equation}

Cytoplasmic cytochrome-{\it c} interacts with Apaf
in an ATP-dependent manner, and the heterodimer forms
a heptameric complex called the ``apoptosome''.\cite{Acehan02} Following
Bagci {\it et al.}\cite{zerrin1}, we model these events via the following
two reactions:
\begin{equation}
\rrxntwoone{cytC}{Apaf}{cytC.Apaf}{40}
\end{equation}
\begin{equation}
\rrxnoneone{7cytC.Apaf}{apop}{41}
\label{e41}
\end{equation}
Further, we use cooperativity in the formation of the apoptosome: the
order of the forward reaction~\ref{e41} is 4.

The apoptosome then incorporates caspase-9, that in its autocatalyzed,
activated form cleaves procaspase-3 (C3)
to form active caspase-3 (C3$^*$). Following 
Albeck {\it et al.},\cite{Albeck08} we use the following reactions for
these processes:
\begin{equation}
\rrxntwoone{apop}{C9}{apop.C9}{42}
\label{e42}
\end{equation}
\begin{equation}
\mmrxn{apop.C9}{C3}{apop.C3.C9}{apop.c9}{C3^*}{43}
\label{e43}
\end{equation}

As mentioned above, tBid is an important factor for activation of
the caspase cascade. Caspase-8, activated due to external death
signals, is an important molecule responsible for Bid cleavage.\cite{Li98}
Additionally, active caspase-3 also results in Bid cleavage, resulting
in a positive feedback loop activating apoptosis.\cite{Slee00}
We model the associated reactions following Bagci {\it et. al.}\cite{zerrin1}:
\begin{equation}
\mmrxn{C3^*}{Bid}{C3*.Bid}{C3^*}{tBid}{44}
\label{e44}
\end{equation}
\begin{equation}
\mmrxn{C8}{Bid}{C8.Bid}{C8}{tBid}{45}
\label{e45}
\end{equation}
The concentration of active caspase-8 depends on external death
factors.

XIAP inhibits the activity of apoptosome\cite{Hill04,Twiddy04} and
promotes proteasomal degradation of caspase-3\cite{Suzuki01}
preventing cell death. On the other hand, Smac inhibits the activity
of XIAP.\cite{Du00} These processes are modeled as,\cite{Albeck08}
\begin{equation}
\rrxntwoone{apop.C9}{XIAP}{apop.C9.XIAP}{46}
\label{e46}
\end{equation}
\begin{equation}
\mmrxn{C3^*}{XIAP}{C3^*.XIAP}{C3^*_{Ub}}{XIAP}{47}
\label{e47}
\end{equation}
\begin{equation}
\mmrxn{C9}{XIAP}{C9.XIAP}{C9_{Ub}}{XIAP}{48}
\label{e48}
\end{equation}
\begin{equation}
\rrxntwoone{Smac}{XIAP}{Smac.XIAP}{49}
\label{e49}
\end{equation}

 In addition to above reactions, there are formation reactions for several
species -- Bcl2, Bax, Apaf, C9, C3, XIAP, Smacm, cytcm, C8, and Bid.
Further, these species, along with C3*, C3U, C9U, Bax*, and p53 forms
mentioned above
also undergo degradation reactions to help establish steady state.

\section{Drug interactions}

 In this section, we integrate the possible drugs to regulate apoptosis
in normal cells into the biochemical network discussed above.
In particular, we consider four specific proteins targets --
Mdm2, PUMA, Bid, and C3$^*$ (Mdm2-I, PUMA-I, Bid-I, and C3-I, respectively).
The actions of the inhibitors are modeled by the following reactions:
\begin{equation}
\rrxntwoone{Mdm2}{Mdm2-I}{Mdm2.Mdm2-I}{50},
\label{e50}
\end{equation}
\begin{equation}
\rrxntwoone{PUMA}{PUMA-I}{PUMA.PUMA-I}{51},
\label{e51}
\end{equation}
\begin{equation}
\rrxntwoone{Bid}{Bid-I}{Bid.Bid-I}{52},
\label{e52}
\end{equation}
and
\begin{equation}
\rrxntwoone{C3^*}{C3-I}{C3^*.C3-I}{53}.
\label{e53}
\end{equation}

 Known inhibitors of p53/Mdm2 interaction includes nutlins, BEB55, BEB59,
and BEB69. Mustata {\it et al.}\cite{Mustata10} have identified several
PUMA inhibitors using pharmacophore modeling, and several of these compounds
have been shown to inhibit PUMA--induced apoptosis in vitro.

\section{Stochastic simulation method}

 The number of proteins in a typical cell (about picoliter) are significantly
less than Avogadro's number: a nanomolar concentration of a protein 
corresponds to $\sim 600$ molecules of that protein in the cell.
Accordingly, for smaller cells and/or lower concentrations,
deterministic rate equations are not applicable. For this
reason, we use the well--known Gillespie algorithm for stochastic simulations
of a system of chemical reactions.

 Details of the Gillespie Algorithm are given in the original 
papers.\cite{gillespie1,gillespie2} Here we briefly describe some of
the ideas behind the algorithm. 

For illustration, consider the following schematic reaction,
\begin{equation}
\mathrm{A}+\mathrm{B}\rightarrow\mathrm{AB},
\label{e54}
\end{equation}
and denote the stochastic rate constant by $c$. This stochastic
rate is directly related to the macroscopic kinetic rate, $k$, by a simple
relation (which, for the reaction in eq~\ref{e1} is $c=kV$, where $V$
is the reaction volume). At any instant of time, the propensity for
the reaction in eq~\ref{e52} is
\begin{equation}
a_{\alpha}=N_{\mathrm{A}}N_{\mathrm{B}}c
\label{e55}
\end{equation}
where $N_{\mathrm{A}}$ and $N_{\mathrm{B}}$ are the number of molecules of
A and B in the reaction volume, and the subscript $\alpha$ denotes that
the reaction index is $\alpha$ in the system of $M$ chemical reactions
($1\le\alpha\le M$).

In the Gillespie algorithm, one such reaction is chosen to be proportional to
its propensity (we use the so-called Direct Reaction version), and
the time is advanced based on the overall reaction propensity at
that time.\cite{gillespie1,gillespie2} The underlying assumptions
are that the system is in thermal equilibrium (the number of
reactive collisions are much less than the number of non-reactive
collisions), and that it is well--mixed.

\subsection{Heterogeneity}

Real cellular systems are heterogeneous - {\it e.g.}, mitochondria and
cytoplasm offer vastly different environments (and, there may be
heterogeneities within these compartments themselves). Thus, at
a first glance, the well-mixed assumption of the Gillespie
algorithm seems to preclude an application to real cellular
systems. However, Bernstein\cite{bernstein1} developed
a protocol to extend the Gillespie algorithm to heterogeneous environments.

In that protocol,\cite{bernstein1} the cell (or, any simulation volume) is
subdivided into elements and a reaction is reproduced and treated
as an individual reaction in each subvolume. Accordingly, each
reactant is treated as a different molecule in each subvolume
({\it e.g.}, $\mathrm{A}_i\neq\mathrm{A}_j$, where $i$ and $j$
are different subvolumes), and the reaction propensities are altered
accordingly. Further, each subvolume is well--mixed -- thus
allowing for formulating the system using the Gillespie algorithm.
In context of this work, the nucleus, the cytoplasm, and the mitochondrial
membrane are well-mixed, distinct subvolumes.

This protocol results in an increase in the number of reactions both
due to the fact that each reaction is treated independently in
different subvolumes, as well as
additional ``reactions'' (due to diffusion)
relating to the conversion of A$_i$ to A$_j$, as described in detail by
Bernstein.\cite{bernstein1}
Based on the identity of a
compartment/subvolume, a type of protein may not be present in that compartment
(for example, the apoptosome does not permeate through the
mitochondrial membrane).

\section{Model parameters}
\label{mp}

 We obtain several parameters from experimental results,
and obtain others in the range of previous computational 
studies.\cite{zerrin1,Albeck08}
Here we discuss some known experimental rate/equilibrium constants, and
the initial concentrations of the species involved in the biochemical
reaction network discussed above.

 The dissociation constant, $k_{27}^-/k_{27}^+$, for p53 and pro-apoptotic 
Bcl-2 family members have been given variously as 160~nM\cite{Chipuk05}
and 535~nM\cite{Tomita06} -- both obtained via surface plasmon resonance. 
We choose an intermediate value of the dissociation constant (333~nM).
PUMA binds significantly more strongly, and 
Chipuk {\it et al.}\cite{Chipuk05} report the dissociation constant
$k_{28}^-/k_{28}^+ = 10$ nM, a value that we use.

 Edlich {\it et al.}\cite{Edlich11} report the on and off rate of
Bax from cytosol and mitochondria as approximately the same, and
establish a first order reaction (as represented by reaction~\ref{e3})
with $k_{29}^+=4.9\times 10^{-3}\mbox{ s}^{-1}$ 
and $k_{29}^-=4.7\times 10^{-3}\mbox{ s}^{-1}$.

 For reaction~\ref{e31}, it is reported that p53 displaces Bax from
its complex with Bcl-2 at equimolar concentration, whereas Bax displaces
p53 from its complex with Bcl-2 at a 50 times higher 
concentration.\cite{Chipuk04} This implies that $k_{31}^+/k_{31}^-=50$.
Since eqs~\ref{e27}, \ref{e30}, and \ref{e31} are not independent,
the above stated values imply that 
$k_{30}^-/k_{30}^+\approx 17$~$\mathrm{\mu}$M.
This value of dissociation constant is significantly higher than
values reported in the literature (0.1~$\mathrm{\mu}$M\cite{Fletcher08}
and 20~nM\cite{Ku11}).
We use $k_{30}^-/k_{30}^+$~nM, except where noted explicitly --
and, as shown below, the results obtained are robust to a wide range
of values for this dissociation constant.

 Half lifes of a few proteins have also been reported in vivo. However,
the direct use of such an information (for example, to compute
the decay rates of the type A$\rightarrow \Phi$) is often not
appropriate because the decay rate of protein A in vivo is
convoluted with, for example, the decay of protein B that upregulates A.

The parameters for the p53 module that differ for these three levels 
of radiation are given in Table~\ref{t3}, and the parameters that stay the
same irrespective of the radiation level are given in Table~\ref{t4}.
These parameters for the p53 module are chosen to reflect p53 oscillations
observed experimentally. $\kappa_{11}$ and $\kappa_{12}$ do not affect the
dose-dependent oscillatory behavior of p53, however, they are important
for coupling of the p53 module to the apoptotic network, and the
corresponding values are given in the text and figures below.

\subsection{Initial concentrations}

 When a normally unstressed cell is exposed to radiation, it
undergoes a sequence of biochemical events that may, depending upon
the radiation dose, lead to apoptosis. Thus, the appropriate initial conditions
of cellular protein levels for such an apoptotic study are the steady
state conditions in an unstressed cell.

 Irrespective of what initial number of molecules we start with,
we first allow the system to establish a normal steady state (without
radiation--induced apoptosis) by selecting N-IR kinetic parameters
from Table~\ref{t3} (in addition to radiation--independent kinetic 
equations/parameters). Subsequently, we perturb the system
with radiation damage (modeled here as genotoxic effect), and allow
the system to evolve. With enough stress, the system will eventually
display apoptosis, with a high concentration of pro-apoptotic
proteins (such as active caspase-3). In essence, the initial concentrations
that are used to study radiation induced damage and subsequent drug
treatments are the steady state conditions without any
radiation/drug.

 For reference, the typical concentration a typical protein present in the cell
under normal conditions is 10--100 nm.\cite{Albeck08} In a picoliter
cell, the number of molecules of a typical protein is, thus,
in the range 6000--60000. On the other hand, proteins such as active caspase-3
(C3$^*$) are expected to be low under normal homeostatic conditions.

\subsection{Modeling radiation damage}

 Mitochondrial apoptosis is regulated by transcription--dependent
and independent role of p53. Thus, we focus on
the p53 module (Figure~\ref{f0}) to model the effects of
radiation damage. The downstream action of the apoptotic pathways
(for example, represented by Bax oligomerization, cleavage of caspase-3) occur
due to the response of the p53 module to radiation damage.
Specifically, we model the response of p53 to three different
radiation levels: no radiation (N-IR), low radiation (L-IR), and
high radiation dose (H-IR).

 The procedure for the simulation is as follows. (1) Establish steady
state for an unstressed cell (N-IR). (2) induce radiation damage
(L-IR or H-IR), as modeled by use of the appropriate kinetic
parameters from Table~\ref{t3} for a certain time $\Delta T$. (3) Return to
unstressed parameters after $\Delta T$.

 We note here that $\Delta T$ {\it does not} refer to the time
duration of radiation exposure, rather it refers to the amount of time
the radiation modifies the kinetic parameters. $\Delta T$ only indirectly
refers to the exposure time: longer exposure to a given radiation
dose will affect cellular biochemical kinetics for a longer duration.

\section{Results}

\subsection{Radiation induced p53 oscillations}

 Single--cell experiments to monitor cellular p53 levels after radiation
damage showed that individual cells show dramatic oscillations in p53
levels. In particular, these oscillations are sustained and show
dose-dependent oscillation frequencies.\cite{Zatorsky06}
These experimentally observed results form the basis of simulating
the p53 module using stochastic simulations: we reproduce the
radiation dependent behavior of p53 response using stochastic simulations.

 Figure~\ref{f1} shows the oscillations in p53n (the major p53 component upon
radiation damage) as a function of time for two different radiation
doses and exposures. 
The chosen model for radiation damage reproduces the known p53 oscillations,
that are dependent upon the radiation dose. At low levels, this
frequency is $\approx 11$~hr, whereas it decreases to $\approx 6$~hr
upon increasing the radiation dose. 

 The differences in L-IR and H-IR parameters for the p53 module
suggests one possible mechanisms
for the observed radiation-dose dependent frequencies. Important
factors in the model leading to this difference are the enhanced
translocation of Mdm2 protein from the cytoplasm to the nucleus, and
the enhanced translocation of Mdm2 mRNA from the nucleus to the
cytoplasm. This was deliberately chosen to represent the time delay
in Mdm2 response that is frequently utilized in deterministic
models.\cite{Ma05,Wagner08,Bottani07}

Thus, this study shows that such time delay can be modeled
mechanistically in stochastic simulations (and allow for the modulation
of oscillation frequency) by the realization that
proteins/genes act in specific cellular contexts. Further, the
increase in the translocation rate of the Mdm protein (along with increased
translocation of p53) to the nucleus upon DNA damage has been 
suggested.\cite{Li02}

Further, as Figure~\ref{f1} suggests,
at low exposure time (modeled by 
$\Delta T=2\times 10^4$~s), there is a single peak in the p53 profile for
both radiation levels. On the other hand, sustained p53 oscillations
are obtained for high exposure times ($\Delta T=2\times 10^5$s).
Even for cells exposed to radiation for the same amount of time,
it is possible that cell--cell variability can lead to different
amounts of observed oscillatory peaks ($\Delta T$ is, after all, not
the radiation exposure time, but is the time for which the kinetic
parameters are altered from unstressed levels).

\subsection{Role of p53 oscillations in apoptosis}

 At higher radiation dosage, the oscillations in p53 occur at a
higher frequency. However, it is unclear if cellular apoptotic
response is mainly determined by the oscillation frequency: is
increased apoptosis at higher dose of radiation a direct result
of the higher p53 oscillation frequency?

 To address this question, we compare the caspase 3 activity
obtained via the use of H-IR and L-IR parameters in Table~\ref{t3},
along with the kinetic parameters associated with the rest of the
apoptotic machinery. Further, we use the same rates for
apoptosis-related transcriptional activation of p53 
($\kappa_{12}=10^{-5}$~s$^{-1}$),
and for p53 translocation to mitochondrial membrane
($\kappa_{11}=10^{-5}$~s$^{-1}$).
The obtained caspase-3 concentrations for a cell are shown in 
Figure~\ref{f4} for both H-IR and L-IR sets.

 Upon using a radiation-dose independent coupling of the p53 module 
with the apoptotic network, L-IR leads to an increase in caspase-3
activity upon radiation damage (radiation damage occurs at $t=0$).
On the other hand, H-IR leads to low caspase-3 concentration.
Clearly, this result is at odds with experimental observation --
increased radiation dose leads to an increase in apoptosis.
Correspondingly, if $\kappa_{11}$ and/or $\kappa_{12}$ are
increased for H-IR, caspase-3 activity is increased (shown
in Figure~\ref{f4}).

 A crucial insight that emerges from Figure~\ref{f4} is that 
the dose-dependent radiation
damage and cell death depend upon the {\it coupling} of the regulatory
p53 module to the apoptotic machinery (defined via an increase
in the p53 pro-apoptotic transcriptional activity, $\kappa_{12}$,
and an increase in p53 translocation to the mitochondria, $\kappa_{11}$), 
and not merely on the p53-oscillation frequencies.
We use the term ``coupling'' to describe how the p53-regulatory module
leads to apoptosis via transcription-dependent ($\kappa_{12}$) and
transcription-independent ($\kappa_{11}$) pathways.

\subsection{Downstream apoptosis cascade}

 The response of the apoptotic network downstream of the p53 module
sheds further light on the role of pro- and anti-apoptotic proteins
in the cell. Here, we discuss (i) oscillations downstream of
p53, (ii) role of Bax--Bcl-2 interactions, and
(iii) the requirement for sufficient Bid cleavage for sustained
apoptosis. 

Figure~\ref{f5} shows the concentrations of PUMA, Bax$^*$, tBid, and caspase-3
after radiation damage with high dose (the qualitative response of
low radiation dose is similar).

PUMA mirrors the p53 oscillations, irrespective of the coupling strength. 
This phenomenon is due to direct activation of
PUMA by p53T. 
Unlike oscillations in PUMA level, oscillations in the concentration of 
activated Bax show a distinct dependence on the coupling strength: at low
coupling strength (no apoptosis), Bax oscillations are sustained, but
decay along with PUMA oscillations. On the other hand, a high coupling
that results in apoptosis leads to a sustained increase in activated Bax
at the expense of oscillations.
Bax$^*$ increases initially due
to the direct action of PUMA on Bax. This leads to the release of 
caspase-3 that cleaves tBid resulting in further Bax activation and 
a positive feedback loop. 

A sustained caspase-3 activity even after the decay of PUMA depends
upon whether sufficient tBid is activated by the time PUMA
decays. A sustained release of tBid and the
positive feedback is very important for sustained caspase activity: for
low coupling of p53 module and apoptotic network, insufficient PUMA activity
results in an insufficient amount of
caspase 3 release and tBid activation to maintain a sustained caspase 
activity.

\subsubsection{Strength of interaction between Bax and Bcl-2}

 In Section~\ref{mp}, we discussed the strength of Baxm and Bcl-2
interactions ($k_{30}^-/k_{30}^+$): different values of this
dissociation constant 
(either directly or indirectly) are suggested by 
different groups spanning several orders of 
magnitude.\cite{Fletcher08,Ku11,Chipuk04}
Here, we discuss the effect of a range of this dissociation
constant on the apoptosis model we use in this manuscript.

 Figure~\ref{f8} shows caspase-3 and Baxm.Bcl-2 concentrations
for two different values of the dissociation constants. A
comparison of panels (a) and (c) shows that the downstream apoptotic
activity is unaffected by dissociation constants that differ by
two orders of magnitude (all other model parameters remain unchanged between
these two panels): caspase-3 concentration is unaffected, although
vastly differing amounts of Baxm.Bcl-2 complex are formed.
Similarly, the different dissociation constants had an insignificant
effect on systems showing a lack of apoptosis (panels (b) and (d)).

 The model is, thus, robust with respect to the strength of
Baxm and Bcl-2 interaction. This suggests that cells can
exhibit significantly differing strengths of this interaction without
affecting the cell fate given a genotoxic insult, and several
alternate mechanisms can lead to the same cell fate.

\subsection{p53 transcription-independent apoptosis}

 So far, we have discussed p53 transcription dependent apoptosis
via transcriptional upregulation of PUMA and Bax upon radiation
damage. On the other hand, the role of p53 in a transcriptionally
independent manner is also of importance.
In its transcription independent role, it has been suggested that
p53 in the cytoplasm can either directly activate Bax,
or can indirectly activate Bax by binding to anti-apoptotic Bcl-2
and, thus, preventing the latter's anti-apoptotic action. In this
section, we explore this issue.

 Figure~\ref{f6} shows the evolution of p53m (top panel) and 
caspase-3 (bottom panel) with time upon exposure to H-IR for
different ratios of transcription dependent ($\kappa_{12}$) and 
independent ($\kappa_{11}$) apoptosis pathway strengths.
In the figure, the black line represents the case where the
strength of transcription dependent pathways alone is not
sufficient to cause apoptosis.
Upon an increase in $\kappa_{11}$ (black, red, and blue lines -- in that
order), p53m shows an increasingly upregulated behavior. Correspondingly,
caspase-3 shows a sustained activity with an increase in
p53 transcription independent pathway strength (for the same
$\kappa_{12}$).

 Strikingly, the removal of direct Bax activation by mitochondrial 
p53 (green line)
results in complete absence of apoptosis for the p53
transcription-independent apoptosis, even for a high strength
of transcription independent pathway. This suggests that the
indirect role of p53m in apoptosis by binding Bcl-2 is not
sufficient for sustained apoptosis by the p53-transcription independent mode -- 
direct Bax activation by the mitochondrial p53 appears to be
critical, too.

\subsection{Role of inhibitors of specific proteins}

  From a pharmacological viewpoint, it is important to identify protein 
targets that are good candidates for drug treatment to mitigate radiation 
damage. In this section, we focus on this issue with the aim of
identifying potential targets suitable for treatments that are effective
even if not administered immediately.

 Figure~\ref{f7} illustrates the efficacy of treatments
administered after two different delays: treatments administered immediately
after radiation damage (15 minute delay, top panel), and treatments
administered after a longer delay (12 hours, bottom panel). The radiation
dose in Figure~\ref{f7} leads to sustained caspase-3 activity in
absence of drug treatment (bottom panel of Figure~\ref{f5}, blue curve).

If treatment is available immediately after radiation damage, 
inhibition of PUMA is very effective in mitigating radiation damage. Similarly,
inhibition of Bid also helps in mitigating radiation damage (although to
a lesser extent than PUMA inhibition).  In contrast, treatment via 
inhibition of Mdm2 and caspase-3 are not effective.
On the other hand, if there is a longer delay before treatment is
administered, PUMA inhibition is an ineffective treatment and inhibition 
of Bid is the most effective treatment. The ineffectiveness of
PUMA inhibition in this case results because PUMA activity till 
the administration of PUMA-I leads to sufficient activation of
positive feedback loop involving caspase-3 and Bid: activation of
Bax via tBid dominates the apoptotic response at this late stage.

 The main cause of apoptosis for the system in Figure~\ref{f7} is
the p53-transcription dependent pathway. Thus, Mdm2 inhibition,
that decreases the ubiquitination of p53n and, subsequently, leads
to an increase in p53T (and PUMA and Bax), increases the
propensity of the system to undergo apoptosis. A comparison
of the two panels of Figure~\ref{f7} shows that administering
Mdm2-I sooner after radiation damage leads to a quicker increase
in caspase-3 activity.

\section{Discussion}

\subsection{Testable hypotheses}

 Several key, testable features of cellular response to apoptosis emerges
from the model and are listed as following. (i) oscillatory response of
PUMA, at the level of a single cell, to radiation damage, (ii) elevated
tBid activity even after PUMA decays when an increase in apoptosis
occurs, and (iii) inhibition of Bid is more effective in mitigating
radiation damage than PUMA inhibition
when the treatment is administered after a substantial delay.

 The last point is especially relevant with respect to developing
a treatment strategy for mitigating radiation damage. Even a very
potent inhibitor of PUMA may not prevent apoptosis when it is
administered after a substantial delay. Indeed, the inhibitor used
in the current model acts in nM concentrations, and despite being
effective when administered immediately after radiation damage, is
ineffective after a longer delay because of enough tBid activity to
maintain a sustained caspase activity.

 Even if potent inhibitors of Bid are not currently known, it is
possible to test the last hypothesis above using the following
two approaches. Firstly, a si-RNA knockout of PUMA performed several hours
{\it after} radiation damage should be ineffective in mitigating apoptosis 
due to radiation damage. Secondly, knocking out Bid several hours after
radiation damage should be more effective in mitigating apoptosis
due to radiation damage.

\clearpage

\bfig
\includegraphics[totalheight=4in]{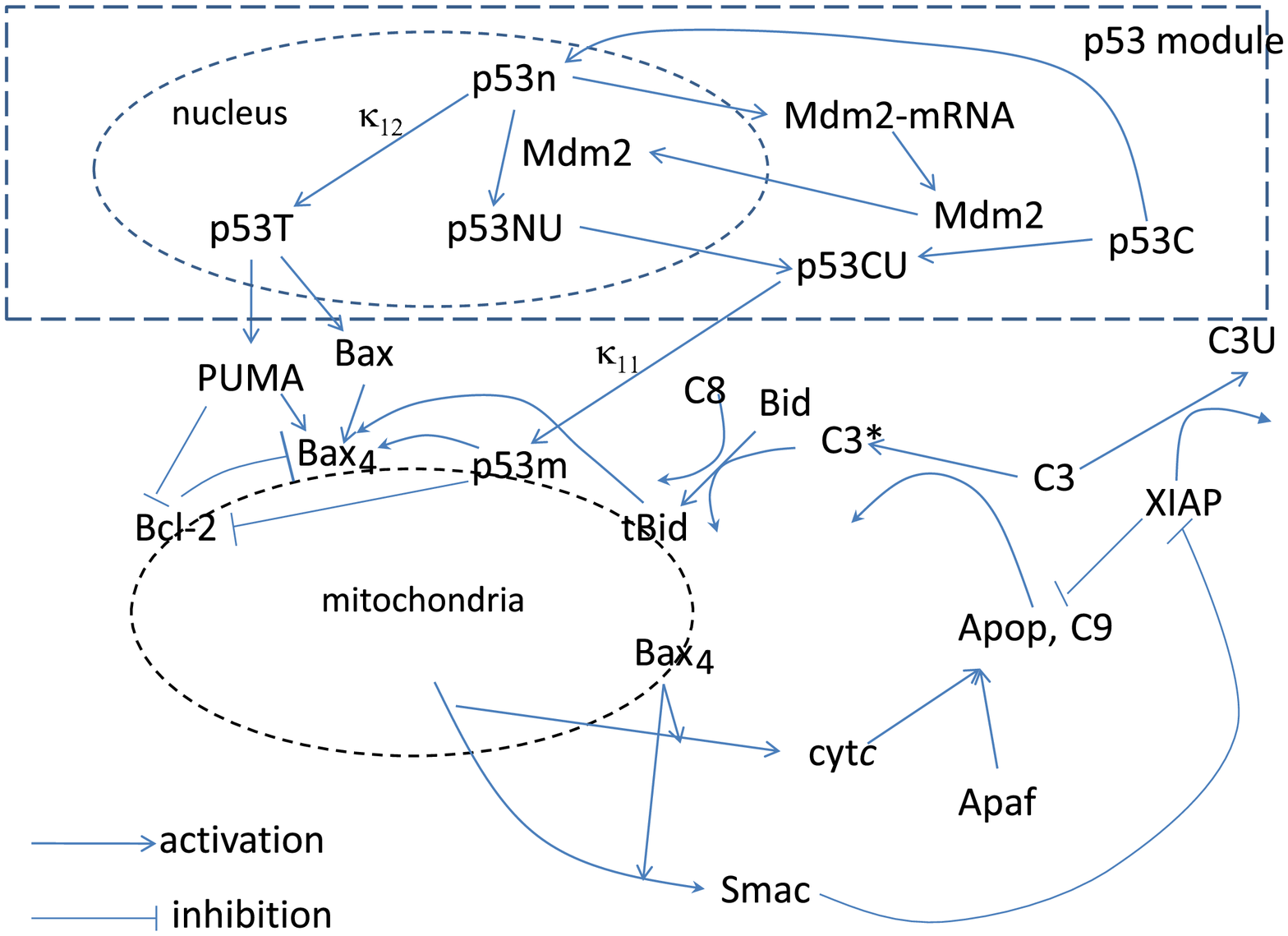}
\caption{A map of interactions in the apoptotic pathway. This map
only shows a schematic of the interactions highlighting the
main features of the model. Detailed biochemical
reactions (that may involve proteins not specifically mentioned in
this map) are given the text.}
\label{f0}
\efig

\clearpage

\bfig
\includegraphics{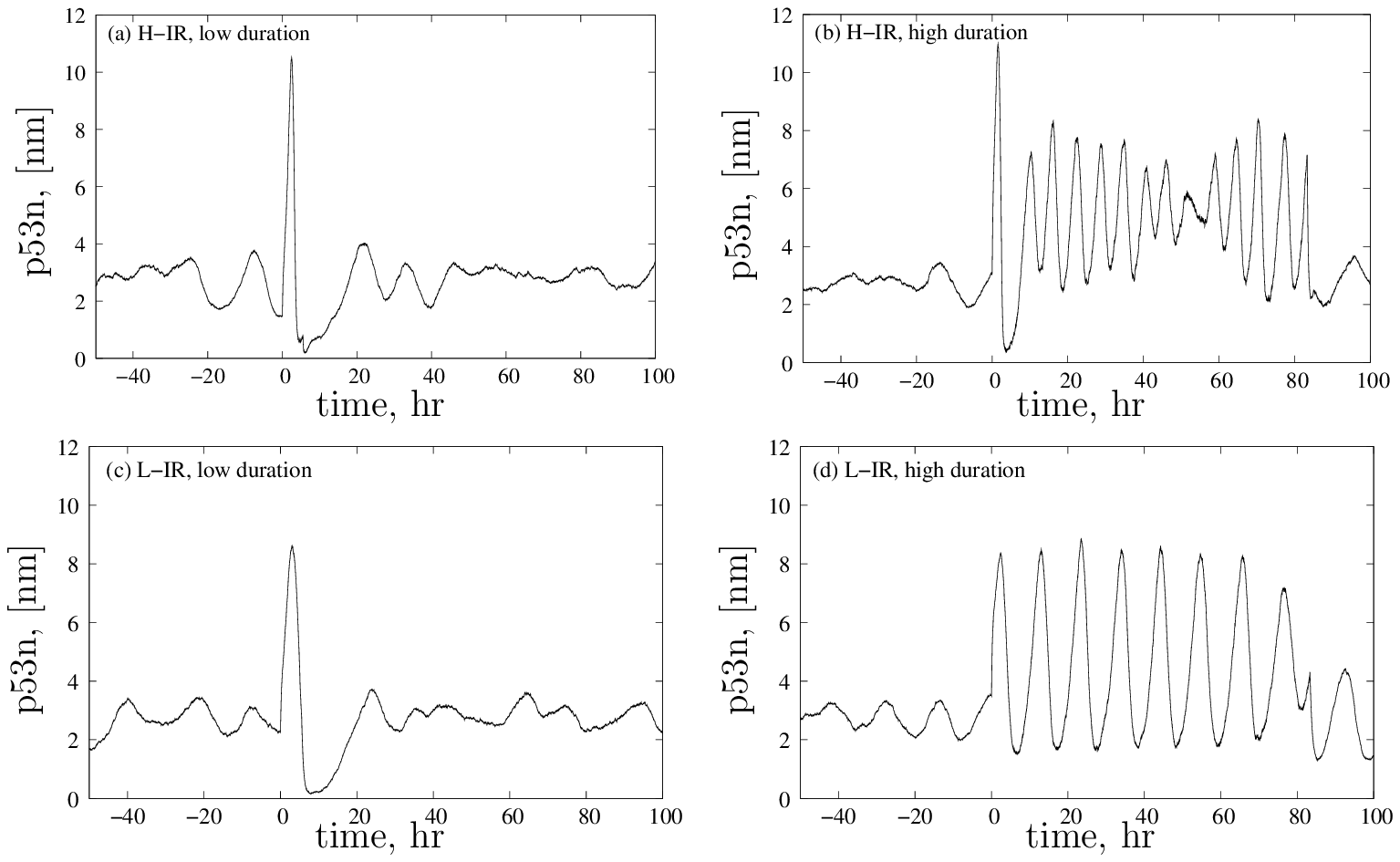}
\caption{Sustained oscillations in p53 concentration as a result of
the duration of alteration, via radiation, of cellular biochemical kinetics.
Radiation damage at $t=0$ leads to radiation--dose dependent p53
oscillations. Single spikes in p53 concentrations (the left two
panels) are observed when
radiation induces a momentary change in cellular biochemical kinetics, and
sustained oscillations (the two panels on the right),
occur when the cellular biochemical kinetics are altered for a longer
duration. Further, the frequencies of oscillations at the two
radiation dosage (top right and bottom right panel) are similar to the
experimentally observed values.}
\label{f1}
\efig

\clearpage

\bfig
\includegraphics{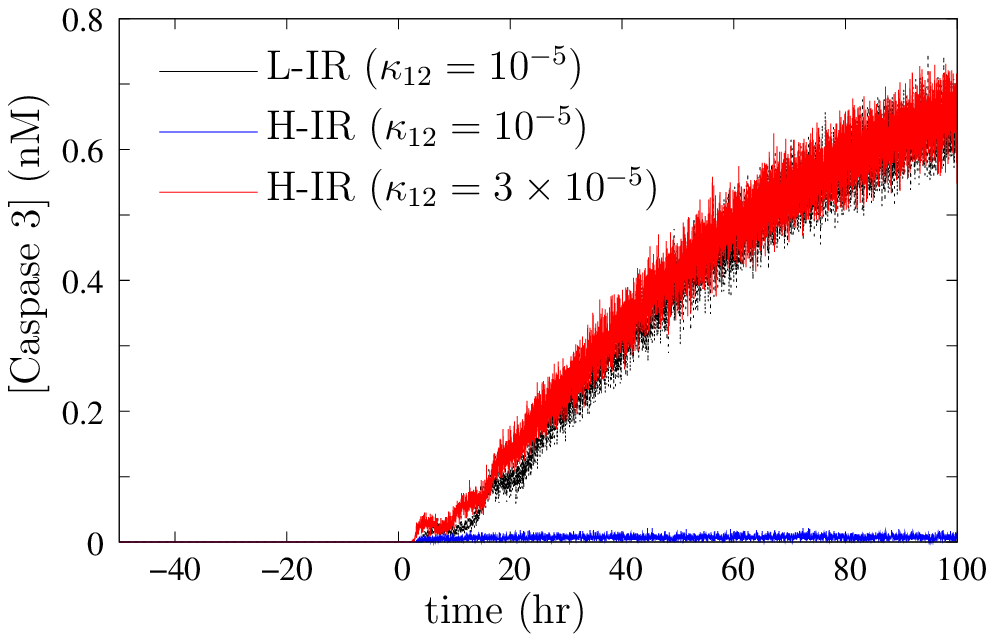}
\caption{An increase in pro-apoptotic transcriptional activity of p53
associates with higher radiation dose to induce apoptosis.
Caspase activities induced by the higher radiation dose for two different
pro-apoptotic transcriptional activity of p53 ($\kappa_{12}$). A lower
dose of radiation leads to a substantial caspase activity at
a smaller value of $\kappa_{12}$.
$\kappa_{11}=10^{-5}$ for the three cases.}
\label{f4}
\efig

\clearpage

\bfig
\includegraphics{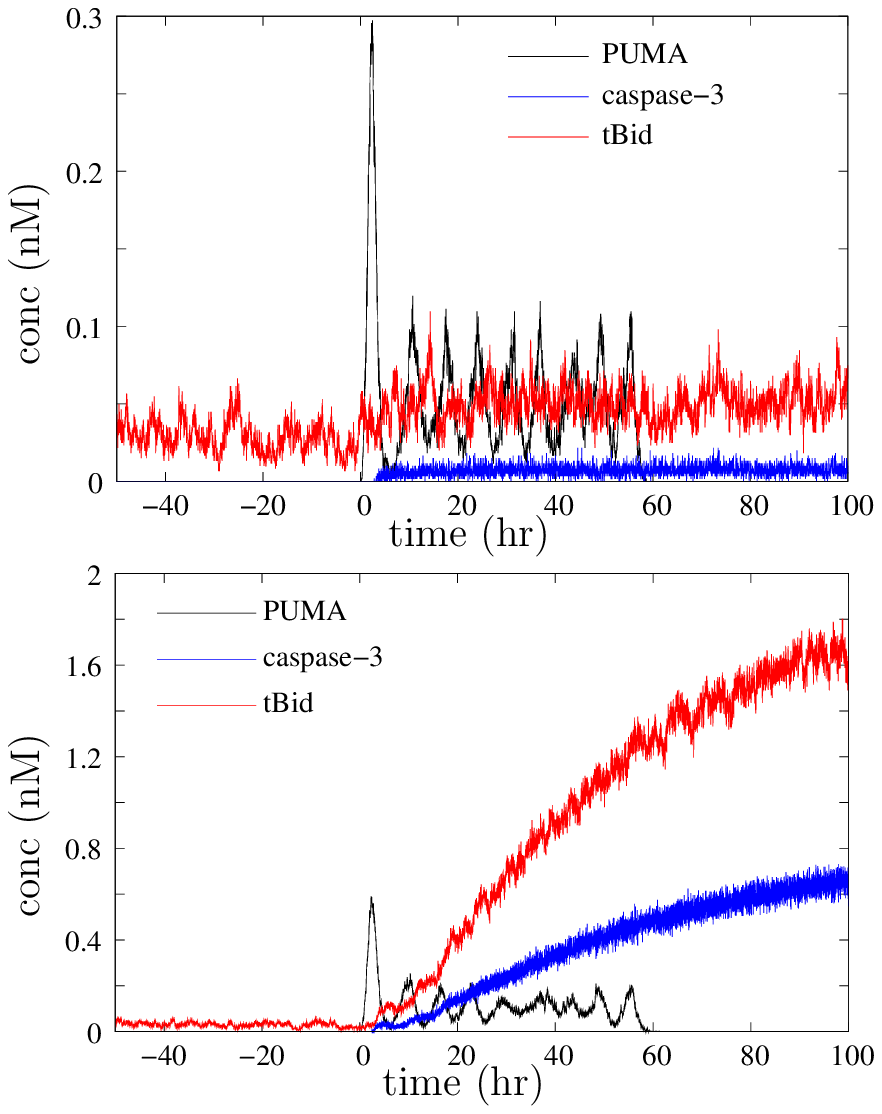}
\caption{Role of tBid activation for sustained caspase-3 activity.
A comparison of the concentrations of PUMA, caspase-3, and tBid
obtained after irradiation with a high dose at $t=0$ with two different
couplings ($\kappa_{12}$) -- low value of $\kappa_{12}$ in the top
panel does not lead to enough tBid activity to sustain apoptotic
activity. $\kappa_{11}=10^{-5}$ for both panels.}
\label{f5}
\efig

\clearpage

\bfig
\includegraphics{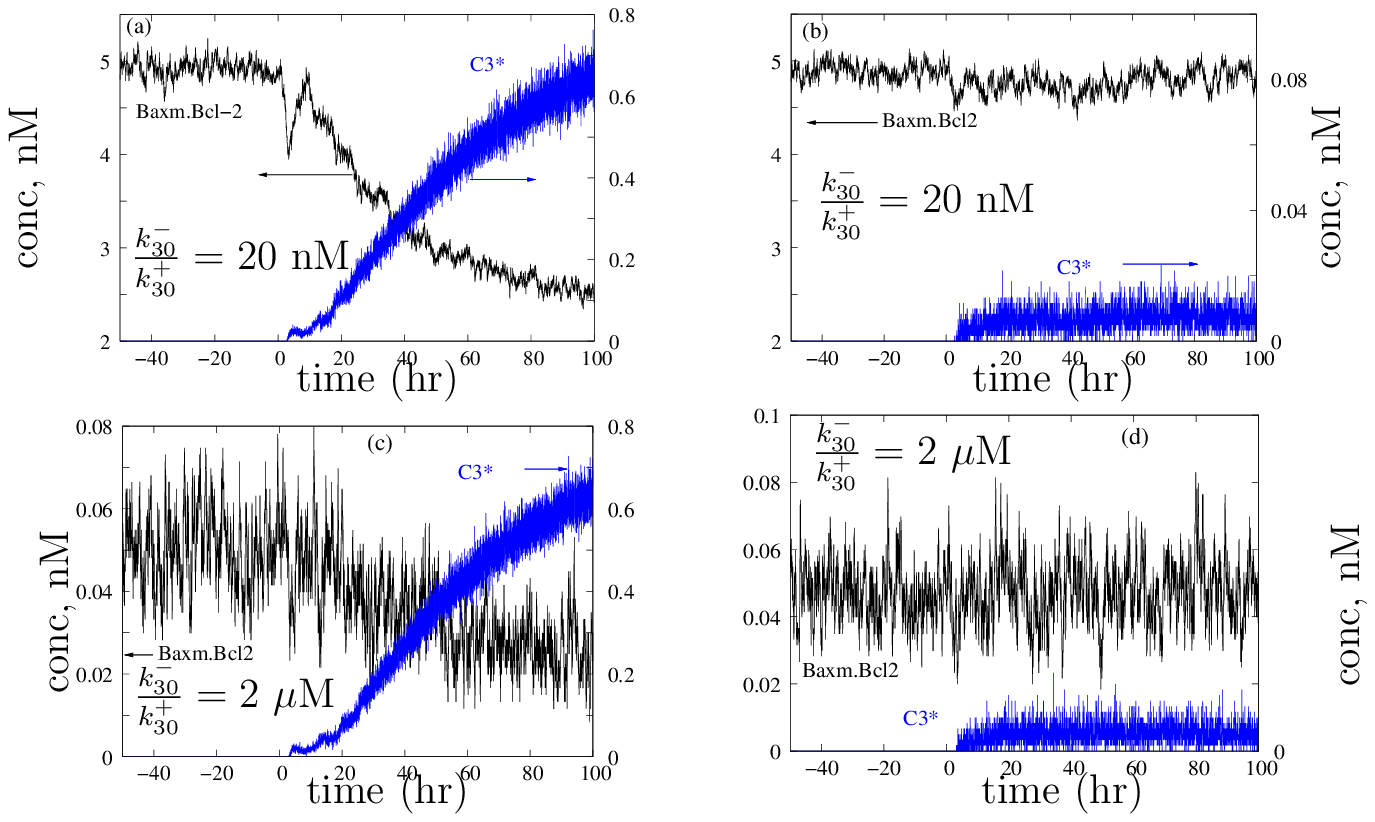}
\caption{Robustness of the model (for caspase-3 activity) with respect to 
the strength of Baxm and Bcl-2 interactions. Panels (a) and (c) show
the results for $\kappa_{12}=2\times 10^{-5}$~s$^{-1}$ showing
distinct apoptosis despite vastly different strengths of Baxm and Bcl-2
interaction. Panels (b) and (d) show an absence of apoptosis for
different strengths of Baxm--Bcl-2 interactions (and both are for
$\kappa_{12}=10^{-5}$~s$^{-1}$.}
\label{f8}
\efig

\clearpage

\bfig
\includegraphics{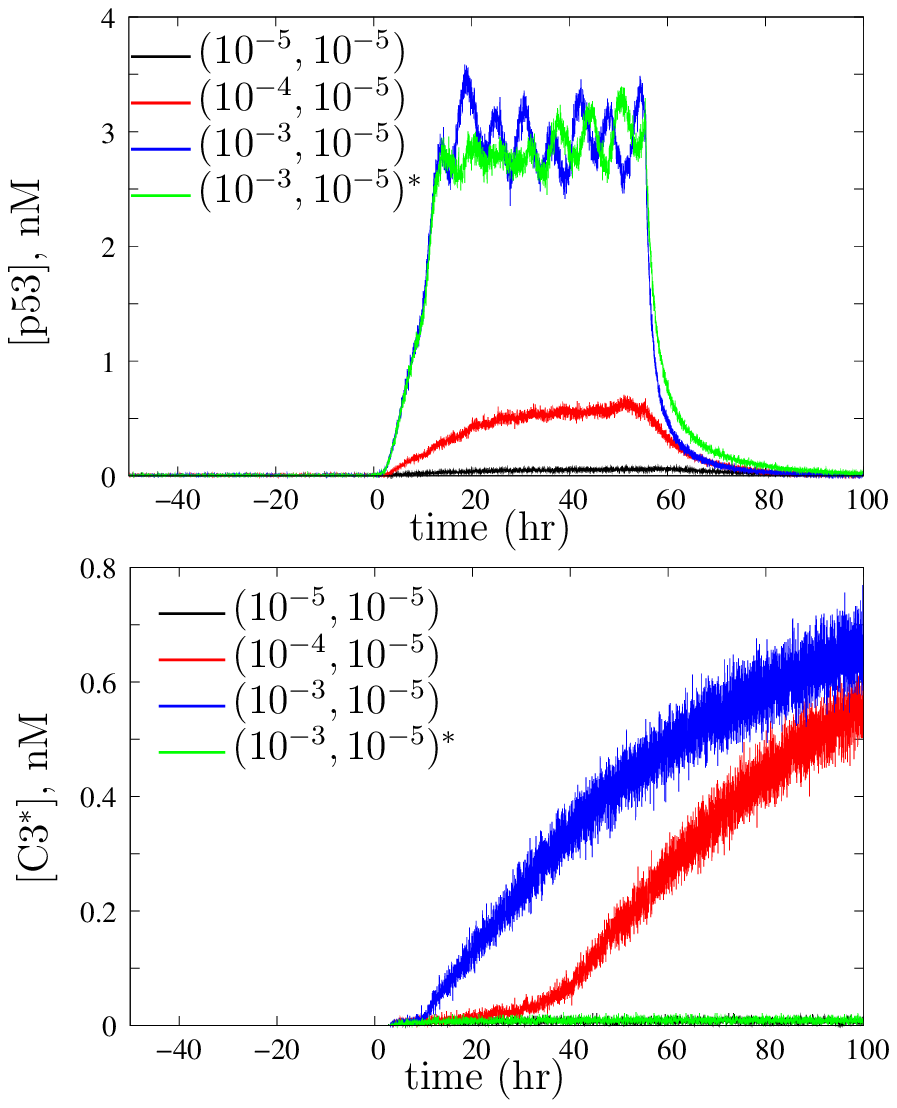}
\caption{Important role of direct Bax activation by mitochondrial p53 in
inducing apoptosis via the p53 transcription-independent pathway.
The top panel shows the mitochondrial p53 concentrations for
different sets of ($\kappa_{11}$,$\kappa_{12}$) values, and the
asterix for the green curve indicates that direct Bax activation by
the mitochondrial p53 in this case is abrogated ($\kappa_{34}=0$). The bottom
panel shows the corresponding C3$^*$ concentrations.}
\label{f6}
\efig

\clearpage

\bfig
\includegraphics{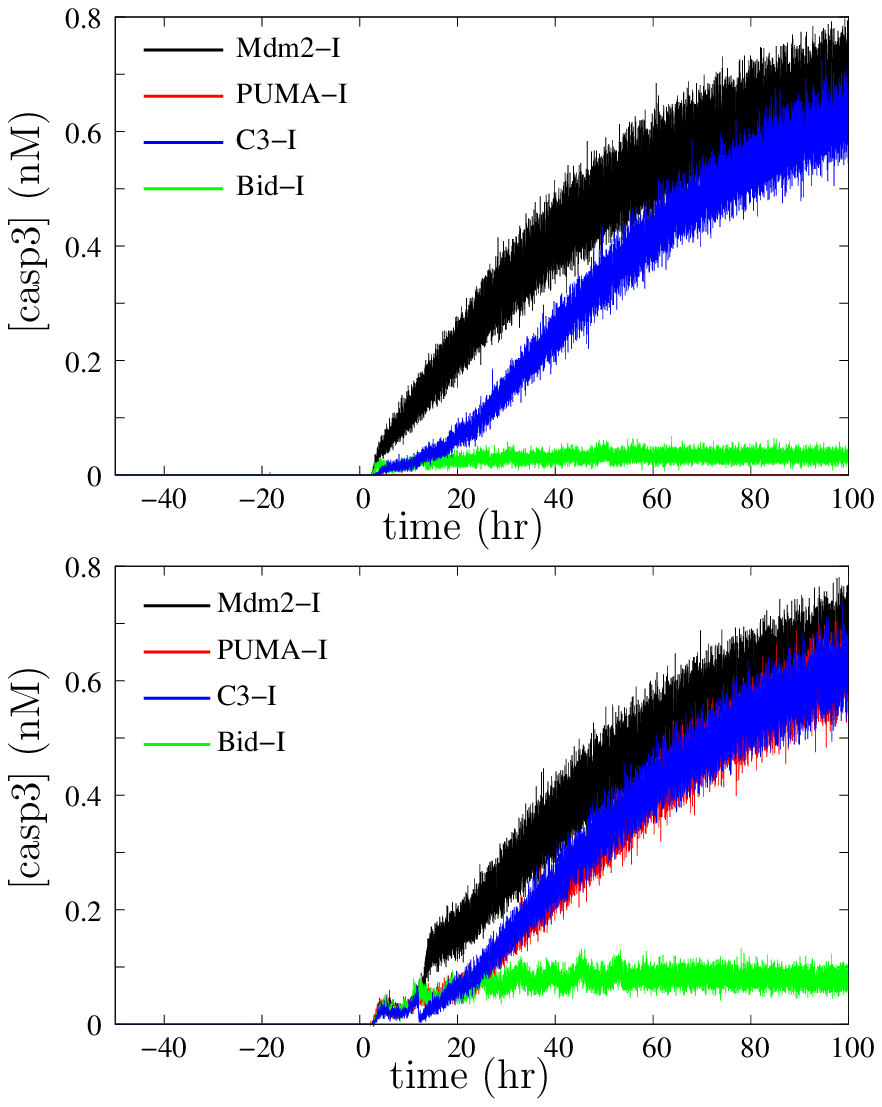}
\caption{Effective mitigation of radiation damage via Bid inhibition at
longer times. The top panel shows caspase-3 activity upon treatment via
Mdm2, PUMA, caspase-3, and Bid inhibition after treatment by the respective 
inhibitors 15 minutes after radiation damage, and the bottom panel shows
caspase-3 activity after treatment administered 12 hours after radiation
damage. In the top panel, caspase-3 activity after treatment with
PUMA inhibitors (PUMA-I) is indistinguishable from zero. In the bottom panel,
resulting caspase-3 activity after treatment by PUMA-I (red line) is similar to
the activity after treatment by procaspase inhibitors (blue line).}
\label{f7}
\efig

\clearpage

\begin{table}
\begin{center}
\caption{List of proteins in Figure~\ref{f0} and their descriptions.}
\begin{tabular}{c|l}
symbol & description \\ \hline
p53n & nuclear p53 \\
p53T & stress--induced p53 for apoptotic activation \\
p53C & p53 in cytoplasm \\
p53CU & ubiquitinated p53 in cytoplasm \\
p53NU & ubiquitinated p53 in nucleus \\
p53U & polyubiquitinated p53 for proteasomal decay \\
p53m & mitochondrial p53 \\
p53no & p53n oligomer for transcription activation of Mdm2 \\
p53To & p53T oligomer for transcription activation pro-apoptotic PUMA/Bax \\
Mdm2-mRNA & mRNA of Mdm2 \\
Mdm2n & Mdm2/E3 ligase proteins in nucleus\\
Mdm2C & Mdm2/E3 ligase proteins in cytoplasm\\
PUMA & p53-upregulated modulator of apoptosis \\
Bcl-2 & anti-apoptotic Bcl-2 family proteins \\
Bax & proapoptotic Bcl-2 family members \\
Bax$_n$ & oligomerized pro-apoptotic Bcl-2 family proteins \\
Baxm & proapoptotic Bcl-2 members on the mitochondria \\
Bid & BH3-only pro-apoptotic proteins \\
tBid & truncated Bid \\
cyt{\it c} & cytoplasmic cytochrome {\it c} \\
Smac & cytoplasmic Smac protein \\
Apaf & Apaf-1 pro-apoptotic protein \\
XIAP & IAP-family proteins \\
apop & apoptosome \\
C9 & caspase 9 \\
C3 & procaspase 3 \\
C3$^*$ & active caspase 3 \\
C3U & ubiquitinated caspase 3 \\
C8 & caspase 8 \\
\end{tabular}
\label{t1}
\end{center}
\end{table}

\clearpage

\begin{table}
\begin{center}
\caption{Radiation dose dependent kinetic parameters for the p53 module.
$\kappa_{11}$ and $\kappa_{12}$ are also dose dependent, but their values
do not affect the observed, dose--dependent p53 oscillatory behavior and are
noted in the text and the figures.}
\begin{tabular}{c|c|c|c}
& No radiation (N-IR) & Low radiation (L-IR) & High radiation (H-IR) \\ \hline
$\kappa_{2}$ (s$^{-1}$) & $10^{-4}$ & $10^{-4}$ & $10^{-3}$ \\
$\kappa_{5}$ (s$^{-1}$) & $2\times 10^{-4}$ & $2\times 10^{-4}$ & $5\times 10^{-3}$ \\
$\kappa_{7}$ (s$^{-1}$) & $10^{-4}$ & $10^{-4}$ & $5\times 10^{-4}$ \\
$k_{15}^+$ ($\mu$M/s) & 0 & $2\times 10^{-7}$ & $5\times 10^{-7}$ \\
$\kappa_{20}$ (s$^{-1}$) & $10^{-4}$ & $10^{-4}$ & $10^{-5}$ \\
$\kappa_{23}$ (s$^{-1}$) & $10^{-4}$ & $10^{-4}$ & $10^{-5}$ \\
\end{tabular}
\label{t3}
\end{center}
\end{table}

\clearpage

\begin{table}
\begin{center}
\caption{Kinetic parameters of the p53 module that are unaffected by
radiation.}
\begin{tabular}{ccc}

\hline

\multicolumn{3}{l}{Formation reactions: $\mathrm{\mu}$Ms$^{-1}$}\\ 
$\kappa_{0}=10^{-6}$ & & \\ \hline

\multicolumn{3}{l}{Unimolecular reactions: s$^{-1}$}\\
$k_{3}^-=10^{-4}$ & $\kappa_{4}=10^{-1}$ & $\kappa_6=5\times 10^{-4}$ \\
$k_{8}^-=0$ & $\kappa_8=10^{-3}$ & $k_{9}^-=0$ \\
$\kappa_9=10^{-3}$ & $\kappa_{10}=10^{-4}$ & $\kappa_{11}=10^{-5}$ \\
$k_{13}^-=10^{-4}$ & $\kappa_{14}=0.1$ & $k_{15}^-=5\times 10^{-5}$ \\
$\kappa_{16}=10^{-2}$ & $\kappa_{17}=0.1$ & $\kappa_{18}=10^{-2}$ \\
$\kappa_{19}=10^{-4}$ & $\kappa_{21}=10^{-3}$ & $\kappa_{22}=10^{-3}$ \\
$\kappa_{24}=10^{-3}$ & & \\ \hline

\multicolumn{3}{l}{Bimolecular reactions: $\mathrm{\mu M}^{-1}$s$^{-1}$}\\ 
$k_8^+=0.1$ & $k_9^+=0.1$ & \\ \hline

\multicolumn{3}{l}{Special (cooperative) reactions}\\
$k_{3}^+=10$ $\mathrm{\mu M}^{-3}$s$^{-1}$ & $k_{13}^+=10$ $\mathrm{\mu M}^3$s$^{-1}$ & \\ \hline

\end{tabular}
\label{t4}
\end{center}
\end{table}

\clearpage

\begin{table}
\begin{center}
\caption{Kinetic parameters downstream of the p53 module.}
\begin{tabular}{ccc}

\hline

\multicolumn{3}{l}{Formation reactions: $\mathrm{\mu}$Ms$^{-1}$}\\ 
\multicolumn{3}{l}{$10^{-5}$ for Bax, Bcl-2, Bid}\\
\multicolumn{3}{l}{$10^{-4}$ for Apaf, C9, C3, XIAP, Smacm, cyt{\it c}}\\ \hline

\multicolumn{3}{l}{Unimolecular reactions: units s$^{-1}$}\\
$\kappa_{25}=10^{-3}$ & $\kappa_{26}=10^{-3}$ & $k_{27}^-=10^{-3}$ \\
$k_{28}^-=10^{-3}$ & $k_{29}^+=5\times 10^{-3}$ & $k_{29}^-=5\times 10^{-3}$ \\
$k_{30}^-=10^{-3}$ & $k_{32}^+=5\times 10^{-3}$ & $k_{32}^-=5\times 10^{-3}$ \\
$k_{36}^-=10^{-3}$ & $k_{37}^-=10^{-3}$ & $k_{40}^-=10^{-3}$ \\
$k_{41}^-=10^{-5}$ & $k_{42}^-=10^{-4}$ & $k_{43}^-=10^{-4}$ \\
$\kappa_{43}=1$ & $k_{44}^-=10^{-3}$ & $\kappa_{44}=1$ \\
$k_{45}^-=10^{-3}$ & $\kappa_{45}=1$ & $k_{46}^-=10^{-4}$ \\
$k_{47}^-=10^{-3}$ & $\kappa_{47}=0.1$ & $k_{48}^-=10^{-3}$ \\
$\kappa_{48}=0.1$ & $k_{49}^-=10^{-3}$ & \\
\multicolumn{3}{l}{$10^{-3}$: PUMA, tBid, Bid, C3, C9, XIAP, Apaf}\\
\multicolumn{3}{l}{$10^{-3}$: cytcm, Smacm, Smac, C3$^*$, C3U, C9U}\\
\multicolumn{3}{l}{$10^{-3}$: Bax, Bax$^*$, Bcl2}\\
\multicolumn{3}{l}{$5\times 10^{-4}$: Mdm2, Mdm2C}\\
\multicolumn{3}{l}{$10^{-2}$: cyt{\it C}}\\
\hline

\multicolumn{3}{l}{Bimolecular reactions: $\mathrm{\mu M}^{-1}$s$^{-1}$}\\ 
$k_{27}^+=3\times 10^{-3}$ & $k_{28}^+=0.1$ & $k_{30}^+=0.6$ \\
$k_{31}^+=1.0$ & $k_{31}^-=2\times 10^{-2}$ & $\kappa_{33}=0.5$ \\
$\kappa_{34}=5\times 10^{-2}$ & $\kappa_{35}=0.5$ & $k_{36}^+=0.6$ \\
$k_{37}^+=0.1$ & $\kappa_{38}=10$ & $\kappa_{39}=10$ \\
$k_{40}^+=0.3$ & $k_{42}^+=3\times 10^{-2}$ & $k_{43}^+=3\times 10^{-2}$ \\
$k_{44}^+=0.3$ & $k_{45}^+=0.3$ & $k_{46}^+=3\times 10^{-2}$ \\
$k_{47}^+=1$ & $k_{48}^+=1$ & $k_{49}^+=0.1$ \\
\hline

\multicolumn{3}{l}{Special (cooperative) reactions}\\
$k_{41}^+=10^5$ $\mathrm{\mu M}^{-3}3$s$^{-1}$ & & \\ \hline

\end{tabular}
\label{t5}
\end{center}
\end{table}

\end{document}